\documentclass[aps,prd,preprint,floatfix]{revtex4}
\usepackage{graphicx}
\def\beq{\begin{eqnarray}}
\def\eeq{\end{eqnarray}}

\begin{document}
\title{Improved Lindstedt-Poincar\'e method for the solution of nonlinear problems}
\author{Paolo Amore}
\email{paolo@ucol.mx}
\author{Alfredo Aranda}
\email{fefo@cgic.ucol.mx}
\affiliation{Facultad de Ciencias, Universidad de Colima, Colima, Mexico}

\begin{abstract}    

We apply the Linear Delta Expansion (LDE) to the Lindstedt-Poincar\'e (``distorted time'') method to 
find improved approximate solutions to nonlinear problems. We find that our method works very well 
for a wide range of parameters in the case of the anharmonic oscillator (Duffing equation) and 
of the non-linear pendulum. The approximate solutions found with this method are better behaved and 
converge more rapidly to the exact ones than in the simple Lindstedt-Poincar\'e method.

\end{abstract}                                                

\maketitle


\section{Introduction}
\label{sec:intro}
The study of nonlinear problems is of crucial importance in all areas of Physics. Some of the
most interesting features of physical systems are hidden in their nonlinear behavior, and can only
be studied with appropriate methods designed to tackle nonlinear problems. In general, given the
nature of nonlinear phenomena, the approximation methods can only be applied within certain ranges 
of the physical parameters and or to only certain classes of problems. It is a challenge to devise 
nonlinear frameworks that contain both operational ease and flexibility in their application.
In this paper we present a method for the solution of nonlinear problems that attempts to 
accomplish these features.

There are several methods which have been used to find approximate solutions to nonlinear problems.
Here we just review a few. Lindstedt developed a method long time ago~\cite{Lin83} in which
one considers solutions to problems involving conservative oscillatory systems with an unknown period. 
The main observation is that by introducing a rescaled time, one can avoid the appearance of terms 
indefinitely growing with time (``secular terms''), that are common in ordinary perturbation theory. 
The method is now know as the Lindstedt-Poincar\'e (LP) method or as the Distorted Time method.

Another known technique is the perturbative $\delta$ expansion (see for example~\cite{BMPS89}). 
In this case the idea is to 
modify the exponent of the nonlinear term by introducing a parameter $\delta$ as new exponent.
$\delta$ interpolates between the linear ($\delta = 0$) and the nonlinear ($\delta=1$) problems. 
If one is able to solve the linear problem then the original nonlinear problem becomes, 
after a power expansion in $\delta$, an infinite sequence of linear 
problems which are (formally) solvable. 

Yet another framework is the Multiple-Scale Perturbation Theory (MSPT)~\cite{BB96}. In this case, one tackles problems
in which a dynamical system has physical behaviors at various length or time scales. This is usually
problematic for ordinary perturbation theory due to the appearance (again) of secular terms. The central idea
is to introduce more than one time and to treat them as independent variables. By performing the usual
perturbative expansion, one then imposes conditions on the solutions (which depend on the different ``times'') 
in order to get rid of secular terms and a linear differential equation is left to solve.

Finally, the Linear Delta Expansion (LDE)~\cite{lde}. This is a method in which an arbitrary 
(or several) parameter $\lambda$
is introduced into the problem and calculations are carried out with conventional perturbation theory
in an expansion parameter $\delta = 1$. At each order in $\delta$, the convergence of the approximation
can be improved by applying the principal of minimal sensitivity which consists on a minimization
of an observable with respect to the parameter $\lambda$.

All of these methods have been applied to a variety of problems. In~\cite{BMPS89}, Bender et al.
showed how one can obtain approximate solutions using the perturbative $\delta$ expansion and
the MSPT to the Duffing  equation (the classical anharmonic  oscillator). Its success then 
has motivated their extension of the method into quantum systems~\cite{BB96}. The LDE 
method has extensively been applied in many different settings with varying degrees of
success. 
For  example, in~\cite{blencowe} it has been used to analyze disordered systems. Pinto and
collaborators have applied it to the Bose-Einstein condensation problem~\cite{Kneur:2002dn}, the
$O(N)(\phi^2)^2_{3d}$ model~\cite{Kneur:2002kq}, to the Walecka model~\cite{Krein:1995rp} 
and to the $\phi^4$ theory at high temperature~\cite{Pinto:1999py}. Detailed references
can be found in these works. 

We can see that it is possible to tackle a large number of nonlinear problems with these
well known techniques. However, there is still room for substantial improvement over them. 
As mentioned before, it is desirable to have a method that works over a large range of parameters,
which is not always the case in the aforementioned methods, and we would like the new 
method to give a smaller error in the approximations than its competitors. 
It is also desirable to devise a framework with operational flexibility and so 
easy to adapt to many different problems.

We show that the method presented in this paper accomplishes these features in the case
of the Duffing equation and of the nonlinear pendulum. The method is based on the application 
of the LDE to the LP method~\cite{AAlet03}. We find solutions that are better behaved and that converge much 
faster than in the other methods described.

In Section~\ref{sec:LP} a brief review of the LP method is presented followed by a
review of the LDE method in Section~\ref{sec:LDE}. We then show the application of both methods to two
problems, the Anharmonic Oscillator in Section~\ref{sec:AO}, and the nonlinear pendulum in
Section~\ref{sec:NP}. We present our conclusions and current work in Section~\ref{sec:conclusions}.
Appendix A contains some of the formulae employed in the 
computations. 

\section{The Lindstedt-Poincar\'e method}
\label{sec:LP}
In this section we introduce the  Lindstedt-Poincar\'e distorted time (LP) method \cite{Lin83}. 
We consider a nonlinear  ODE of the form
\beq
\ddot{x}(t) + \omega^2 \ x(t) = \varepsilon \ f(x(t)) \ ,
\label{LP1}
\eeq
which describes a conservative system, oscillating with an unknown period $T$. The nonlinear term
$\varepsilon \ f(x(t))$ is treated as a perturbation. Unfortunately, when the ordinary perturbation is applied 
to eq. (\ref{LP1}), by writing the solution as a series in $\varepsilon$, the appearance of secular terms spoils 
the expansion and any predictive power is lost for sufficiently large time scales.

In order to avoid the appearance of secular terms, we switch to a scaled time $\tau = 2 \pi t/T \equiv \Omega \ t$, 
where $T$ is the (unknown) period of the oscillations. The ODE now reads:
\beq
\Omega^2 \frac{d^2x}{d\tau^2}(\tau) + \omega^2 \ x(\tau) = \varepsilon \ f(x(\tau)) \ .
\label{LP2}
\eeq

We notice that the dependence upon $\varepsilon$ in this equation enters both in the solution $x(\tau)$ and in the 
frequency $\Omega$. By assuming $\varepsilon$ to be a small parameter we write 
\beq
\Omega^2 &=& \sum_{n=0}^{\infty} \ \varepsilon^n \ \alpha_n \ \ \ ; \ \ \
x(\tau) = \sum_{n=0}^{\infty} \ \varepsilon^n \ x_n(\tau) \nonumber 
\eeq
and expand the r.h.s of eq. (\ref{LP2}) as
\beq
f(x) &=& f\left( \sum_{n=0}^{\infty} \ \varepsilon^n \ x_n(\tau) \right) \approx 
f(x_0) + \varepsilon \ x_1 \ f'(x_0) + \varepsilon^2 \ \left[ x_2 \ f'(x_0) + \frac{x_1^2}{2} \ f''(x_0) \right] \nonumber \\
&+& \varepsilon^3 \ \left[ x_3 \ f'(x_0) +  x_2 \ x_1 \  f''(x_0) + \frac{x_1^3}{6} \ f'''(x_0)
 \right] + O\left[\varepsilon^4\right] \nonumber \ .
\eeq

By using these expansions inside  eq.~(\ref{LP2}) we obtain a system of linear inhomogeneous differential equations, 
each corresponding to a different order in $\varepsilon$. Let us consider the first few terms. 
To order $\varepsilon^0$ we obtain the equation
\beq
\alpha_0 \ \frac{d^2x_0}{d\tau^2} + \omega^2 \ x_0(\tau)  &=& 0 \ ,
\label{LP3}
\eeq
describing a harmonic oscillator of frequency $\Omega = \sqrt{\alpha_0} =\omega$.  
To order $\varepsilon$ we obtain the equation
\beq
\alpha_0 \ \frac{d^2x_1}{d\tau^2} + \omega^2 \ x_1(\tau)  = s_1(\tau) \ ,
\label{LP4}
\eeq
where the r.h.s. is given by
\beq
s_1(\tau) &\equiv& - \alpha_1 \  \frac{d^2x_0}{d\tau^2} + f(x_0) \ .
\eeq

We stress the oscillatory behavior of the driving term $s_1(\tau)$, because of its dependence upon the order-0 solution, $x_0(\tau)$. 
As a result $s_1(\tau)$ will contain the fundamental frequency, corresponding to a period of $2 \pi$ in the scaled time, and 
multiples of this frequency, appearing through the term $f(x_0(\tau))$. The presence of a driving term with the 
fundamental frequency leads to a resonant behavior of $x_1(\tau)$ and to the unfortunate occurrence of secular terms, which
spoils our expansion. However, we can deal with this problem by fixing the coefficient $\alpha_1$ to cancel the resonant 
term in the r.h.s. of eq. (\ref{LP4}). The iteration of this procedure to a given order $n$ allows to determine the coefficients
$\alpha_0, \dots, \alpha_n$ and therefore the frequency $\Omega = \sqrt{\alpha_0 + \epsilon \ \alpha_1 + \dots + \epsilon^n \ \alpha_n}$.

\section{Linear delta expansion}
\label{sec:LDE}

The linear delta expansion (LDE) is a powerful technique which has been originally introduced 
to deal with problems of strong coupling Quantum Field Theory, for which the naive perturbative approach is not 
useful. Since then this method has been applied to a wide class of 
problems~\cite{blencowe,Kneur:2002dn,Kneur:2002kq,Krein:1995rp,Pinto:1999py}.
In its original formulation a lagrangian density ${\cal L}$, which is not exactly solvable, is interpolated with
a solvable lagrangian ${\cal L}_0(\lambda)$, depending upon one (or more) parameters $\lambda$:
\beq
{\cal L}_\delta = {\cal L}_0(\lambda) + \delta \ \left( {\cal L} - {\cal L}_0(\lambda) \right) \ .
\eeq

For $\delta=0$ one obtains ${\cal L}_0(\lambda)$, whereas for  $\delta=1$ one recovers the full lagrangian ${\cal L}_\delta$. 
The term $\delta \ \left( {\cal L} - {\cal L}_0 \right)$ is treated as a perturbation and $\delta$
is used to keep track of the perturbative order. Eventually $\delta$ is set to be $1$. 

We notice that the interpolation of the full lagrangian with the solvable one, ${\cal L}_0(\lambda)$, brings an artificial 
dependence upon the arbitrary parameter $\lambda$. Such dependence, which would vanish if all perturbative orders were calculated,
can be soften to a finite perturbative order, by requiring some physical observable $\cal O$ to be locally insensitive to 
$\lambda$, i.e:
\beq
\frac{\partial{\cal O}(\lambda)}{\partial \lambda} = 0 \nonumber .
\eeq
This condition is known as Principle of Minimal Sensitivity (PMS) and is normally 
seen to improve the convergence to the exact 
solution.

\section{Anharmonic oscillator}
\label{sec:AO}

In this Section we apply the LDE to the LP method in order to find approximate
solutions to the Duffing equation, a problem which has already been considered in \cite{AAlet03};
here we present the calculation in more detail.

Consider the equation for the  anharmonic oscillator
\beq
\frac{d^2x}{dt^2}(t) + \omega^2 \ \ x(t) = - \mu \ x^3(t) \ .
\label{duf1}
\eeq
This equation describes a conservative system, where the total energy is
given by 
\beq
E = \frac{\dot{x}^2}{2} + \left[ \frac{\omega^2 \ x^2}{2} + \mu \ \frac{x^4}{4} \right] \ .
\eeq
The period of the oscillation can be calculated in terms of an elliptic integral
\beq
T_{exact} &=& 2 \ \int_{-A}^{A} dx \ \frac{1}{\sqrt{{2 (E-V(x))}}} \ ,
\label{periodex}
\eeq
where $A$ is the amplitude of the oscillations.

Following the procedure explained in the Section \ref{sec:LP} and \ref{sec:LDE}, we write
Eq.~(\ref{duf1}) as
\beq
\Omega^2 \frac{d^2x}{d\tau^2}(\tau) + \left( \omega^2 + \lambda^2\right) \ x(\tau) = \delta \left[
- \mu \ x^3(\tau) + \lambda^2  \ x(\tau)\right]  \ ,
\label{duf2}
\eeq
where an arbitrary parameter $\lambda$ with dimension of frequency has been introduced. 
Clearly for $\delta = 1$, Eq.~(\ref{duf2}) reduces to Eq.~(\ref{duf1}). We repeat the procedures previously 
explained and find a hierarchy of linear inhomogeneous differential equations 
to be solved sequentially. 

\subsubsection{Zeroth Order}

To zeroth order we obtain the equation
\beq
\alpha_0 \ \frac{d^2x_0}{d\tau^2} + (\omega^2+\lambda^2) \ x_0(\tau)  &=& 0 \ ,
\eeq
with solution
\beq
x_0(\tau) &=& A  \ \cos\tau \ .
\eeq

The zeroth order frequency is then given by
\beq
\alpha_0 &=& \omega^2 +  \lambda^2  \ .
\eeq

\subsubsection{First Order}

To first order we find the equation 
\beq
\alpha_0 \ \frac{d^2x_1}{d\tau^2} + (\omega^2+\lambda^2) \ x_1(\tau)  
          &=& S_1(\tau) \, ,
\eeq
where
\beq
S_1(\tau) &=& A \ \cos\tau 
\left[ \alpha_1 + \lambda^2 - \frac{3 A^2 \mu}{4} \right] - \frac{A^3 \mu}{4} \ \cos 3\tau \ .
\eeq
Now $\alpha_1$ is fixed by eliminating the term proportional to $\cos\tau$:
\beq
\alpha_1 &=& \frac{3 A^2 \mu}{4} - \lambda^2 \ .
\eeq

We obtain the solution 
\beq
x_1(\tau) &=& - \frac{A^3 \ \mu}{ 32 (\omega^2+\lambda^2)} \ \cos\tau+ 
 \frac{A^3 \ \mu}{ 32 (\omega^2+\lambda^2)} \ \cos 3\tau  \nonumber \ ,
\eeq
and the frequency
\beq
\Omega^2 = \alpha_0 +\alpha_1 = \omega^2 + \frac{3 A^2 \mu}{4} \, ,
\eeq
which is observed to be  independent of $\lambda$.

\subsubsection{Second Order}

The second order equation is given by 
\beq
\alpha_0 \ \frac{d^2x_2}{d\tau^2} + (\omega^2+\lambda^2) \ x_2(\tau) 
&=& S_2(\tau) \, ,
\eeq
where now
\beq
S_2(\tau)
&=&
\frac{A \left( 3 \ A^4 \ \mu^2+128 \ \alpha_2 \ (\omega^2+\lambda^2) \right)}{128
 \ (\omega^2+\lambda^2)} \ \cos\tau  \nonumber \\
&+& \frac{A^3 \ \mu \ (3 \ A^2 \ \mu - 4 \ \lambda^2)}{16 \  (\omega^2+\lambda^2) } 
\ \cos 3\tau \\ \nonumber
&-& \frac{3 \ A^5 \ \mu^2}{128 \  (\omega^2+\lambda^2)} \ \cos 5\tau \, .
\eeq
As before $\alpha_2$ is fixed by eliminating the term proportional to $\cos\tau$:
\beq
\alpha_2 &=& - \frac{3 \ A^4 \ \mu^2}{128 \  (\omega^2+\lambda^2)} \, .
\eeq

We obtain the solution
\beq
x_2(\tau) &=& \frac{A^3 \mu \ (23 A^2 \mu-32 \lambda^2)}{1024   (\omega^2+\lambda^2)^2} \ \cos\tau + 
\frac{A^3 \mu (- 3 A^2 \mu+ 4 \lambda^2)}{128  (\omega^2+\lambda^2)^2}  \ \cos 3\tau \nonumber \\
&+& \frac{A^5 \mu^2}{1024  (\omega^2+\lambda^2)^2}  \ \cos 5\tau
\eeq
and the frequency
\beq
\Omega^2 = \alpha_0 +\alpha_1 + \alpha_2 = \omega^2 + \frac{3 A^2 \mu}{4} -
\frac{3 A^4 \mu^2}{128 \  (\omega^2+\lambda^2)}  \ .
\eeq
Note that at this order the frequency now depends on the arbitrary parameter $\lambda$. However, due to the
explicit dependence, by applying the PMS, we would obtain the same solution as in the simple LP method ($\lambda = 0$). 
In order to get a different solution, we must go to the next order in the expansion.

\subsubsection{Third Order}

Following the same procedure, we obtain the following expression for the third order:
\beq
\alpha_0 \ \frac{d^2x_3}{d\tau^2} &+& (\omega^2+\lambda^2) \ x_3(\tau) 
=  S_3(\tau) \, ,
\eeq
where
\beq
s_3(\tau) 
&=& \left[ A \ \alpha_3 - 
\frac{3 \ A^5 \ \mu^2 \ (3 \ A^2 \ \mu - 4 \ \lambda^2)}{512 \ (\omega^2+\lambda^2)^2} \right] \ 
\cos\tau \nonumber \\
&-& \frac{(A^3 \ \mu \ (297 \ A^4 \ \mu^2 - 768 \ A^2 \ \mu \ \lambda^2 + 512 \ 
\lambda^4))}{2048 \ (\omega^2+\lambda^2)^2}  \ 
\cos 3\tau  \nonumber \\ 
&+& \frac{3 \ A^5 \ \mu^2 \ (3 \  A^2 \ \mu-4 \ \lambda^2)}{256 \ (\lambda^2 + \omega^2)^2}
 \ \cos 5\tau - \frac{3 \ A^7 \ \mu^3}{2048 \ (\lambda^2+\omega^2)^2} \ \cos 7\tau \, .
\eeq

By eliminating the term proportional to $\cos\tau$ we determine $\alpha_3$ to be 
\beq
\alpha_3 &=& \frac{3 \ A^4 \ \mu^2 (3 \ A^2 \ \mu - 4 \ \lambda^2)}{512 \ (\lambda^2+\omega^2)^2} \, ,
\eeq
and the solution 
\beq
x_3(\tau) &=& - \frac{A^3 \ \mu}{32768} \ \frac{547 \ A^4 \ \mu^2 - 1472 \ A^2 \ \mu \ \lambda^2 + 1024 \ 
\lambda^4}{(\lambda^2+\omega^2)^3} \ \cos\tau \nonumber \\
&+& \frac{A^3 \ \mu}{16384} \ \frac{297 \ A^4 \ \mu^2 - 768 \ A^2 \ \mu \ \lambda^2 + 512 \ 
\lambda^4}{(\lambda^2+\omega^2)^3} \ \cos 3\tau \nonumber \\
&+& \frac{A^5 \ \mu^2}{2048} \ \frac{(-3 \ A^2 \ \mu + 4 \ \lambda^2)}{(\lambda^2+\omega^2)^3} \ \cos 5\tau 
+ \frac{A^7 \ \mu^3}{32768} \ \frac{1}{(\lambda^2+\omega^2)^3} \ \cos 7\tau \, .
\nonumber 
\eeq

The frequency to order $\delta^3$ is now obtained to be
\beq
\Omega^2 = \alpha_0 +\alpha_1 + \alpha_2 + \alpha_3 = 
\omega^2 + \frac{3 A^2 \mu}{4}  -
\frac{3 \ A^4 \ \mu^2}{128 \  (\omega^2+\lambda^2)} +
\frac{3 \ A^4 \ \mu^2 (3 \ A^2 \ \mu - 4 \lambda^2)}{512 \ (\lambda^2+\omega^2)^2} \, .
\label{omega3duff0}
\eeq
This time, the frequency depends upon the arbitrary parameter $\lambda$ in a nontrivial way and
we can apply the PMS in order to fix the value of $\lambda$. We do this by 
imposing that $\frac{d\Omega^2}{d\lambda} = 0$, which leads to the following result:
\beq
\lambda^2 &=& \frac{3 \ A^2 \ \mu}{4} \, .
\eeq
Notice that since $\lambda$ depends linearly upon $A$ the formula for $\Omega^2$ obtained in this case 
{\sl does not simply correspond to an expansion in $A$}.
As a matter of fact we find that the frequency corresponding to this value of $\lambda$ is
\beq
\Omega^2 = \frac{64 \ A^4 \ \mu^2+192 \ A^2 \ \mu \ \omega^2 + 128 \ \omega^4}{96 \ A^2 \ \mu + 128 \ \omega^2} \, .
\label{omega3duff}
\eeq

Notice that the Duffing equation (\ref{duf1}) is left invariant under the simultaneous rescaling of the anharmonic 
coupling  $\mu$ and of the amplitude, i.e. $\mu \rightarrow \mu'$ and $A \rightarrow A' = A \ \sqrt{\mu/\mu'}$. This
invariance is manifest in the equation (\ref{omega3duff}), which is function of $A^2 \ \mu$, which is invariant under
this rescaling.

\begin{figure}[t]
\includegraphics[width=11cm]{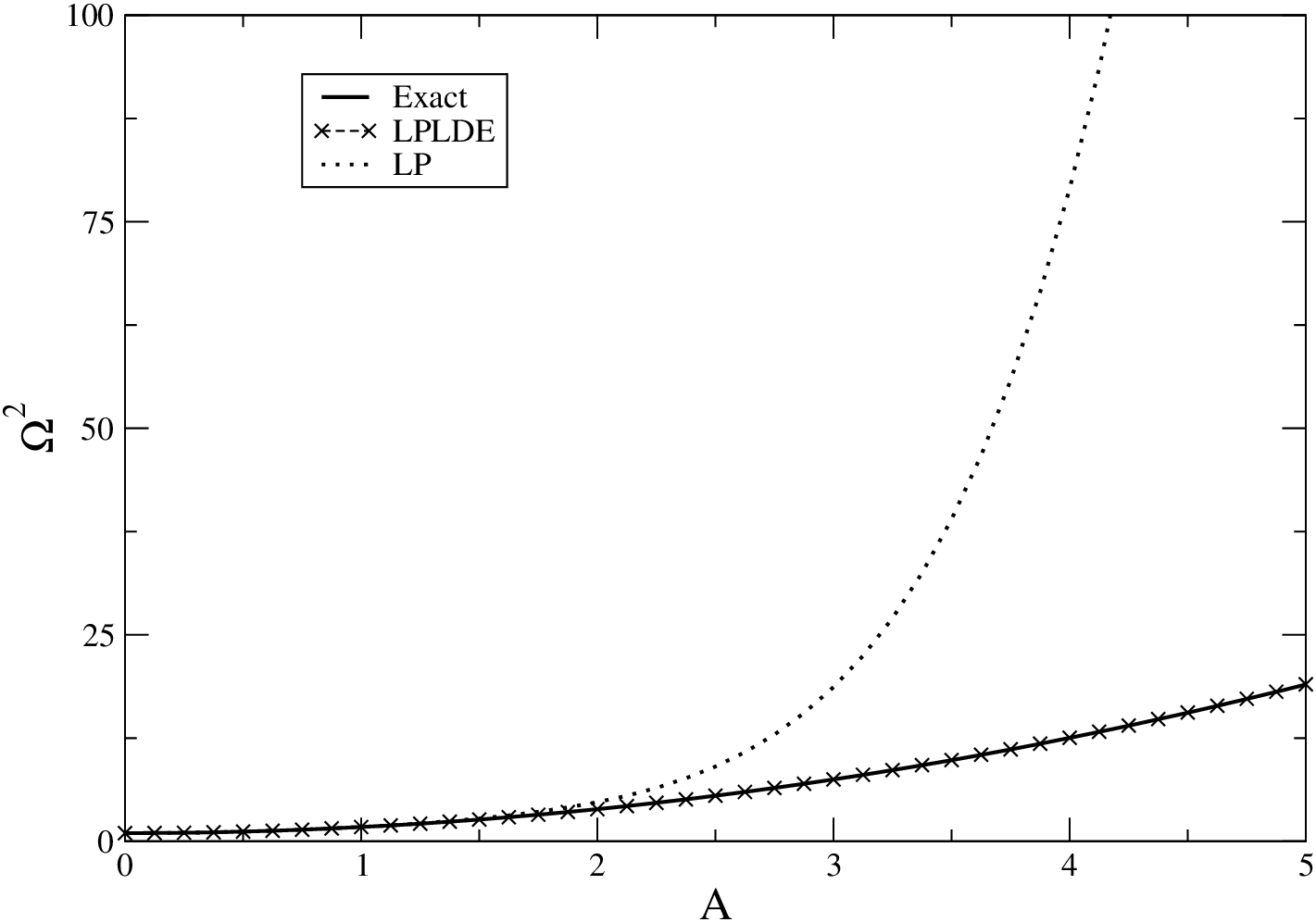}
\caption{Squared frequency of the anharmonic oscillator as a function of the amplitude (arbitrary units). $\omega = \mu =1$. }
\label{fig1}
\end{figure}

\begin{figure}[t]
\includegraphics[width=11cm]{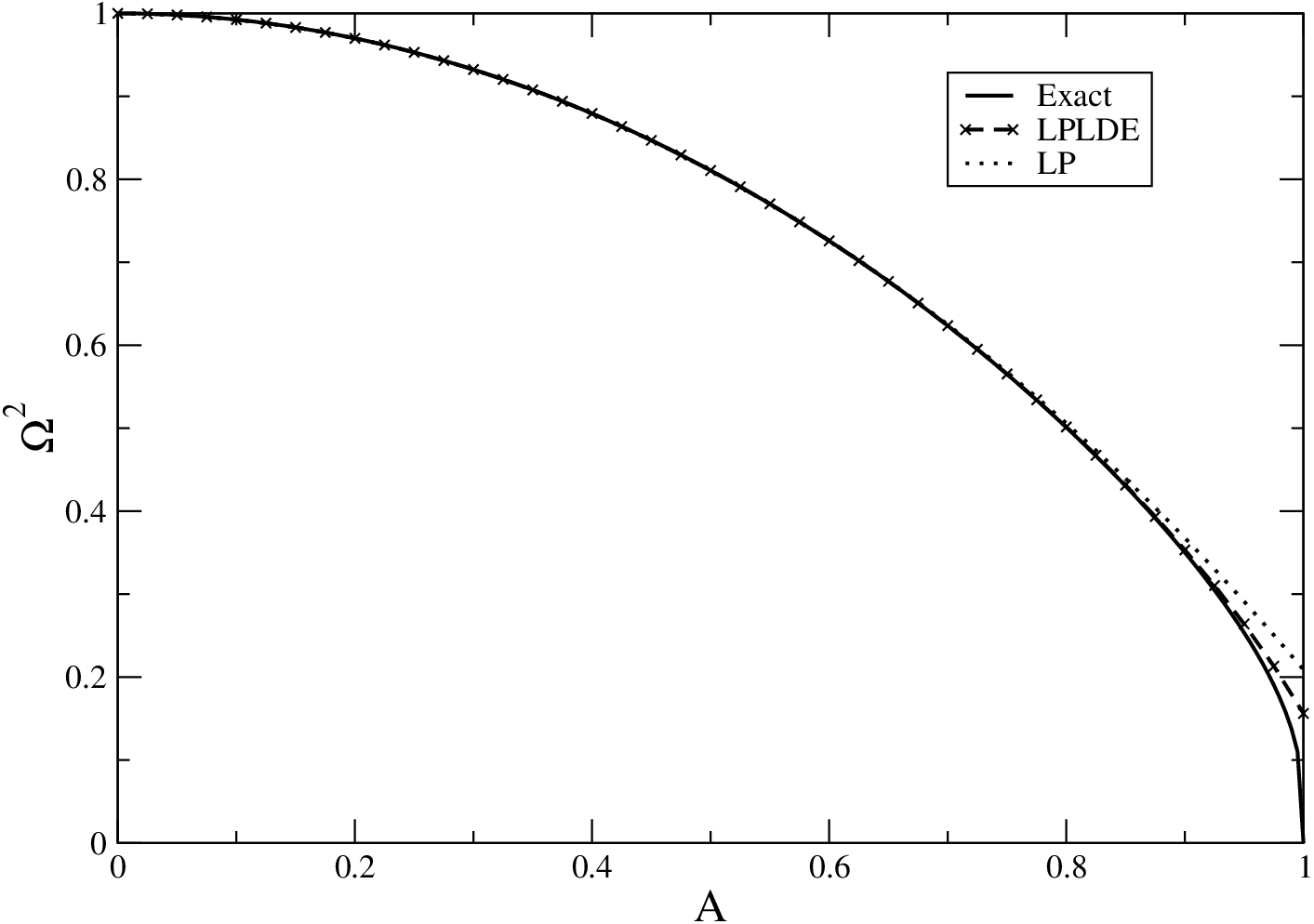}
\caption{Squared frequency of the anharmonic oscillator as a function of the amplitude (arbitrary units). $\omega =1$ and $\mu =-1$. }
\label{fig2}
\end{figure}

\begin{figure}[t]
\includegraphics[height=5.5cm]{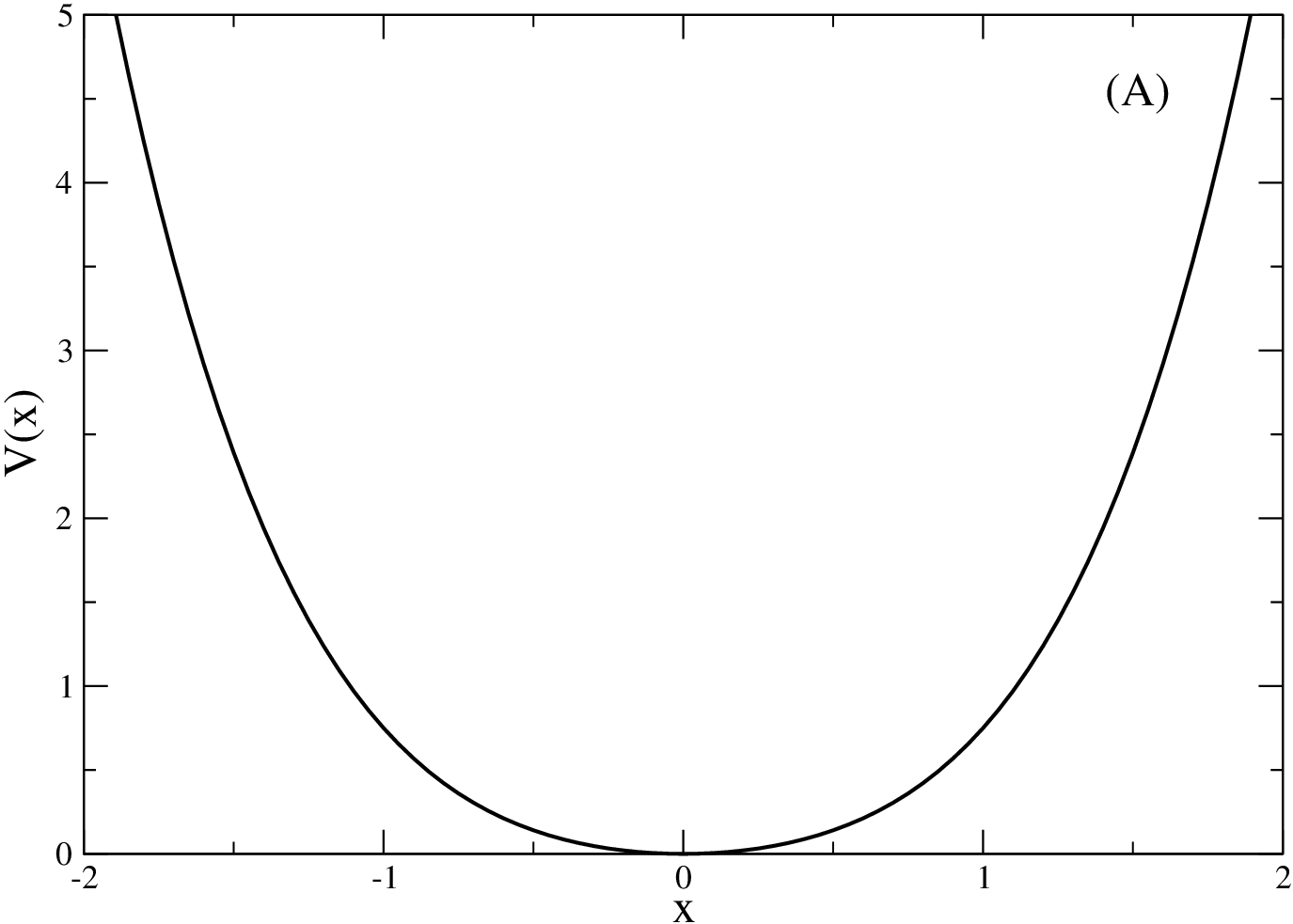}
\includegraphics[height=5.5cm]{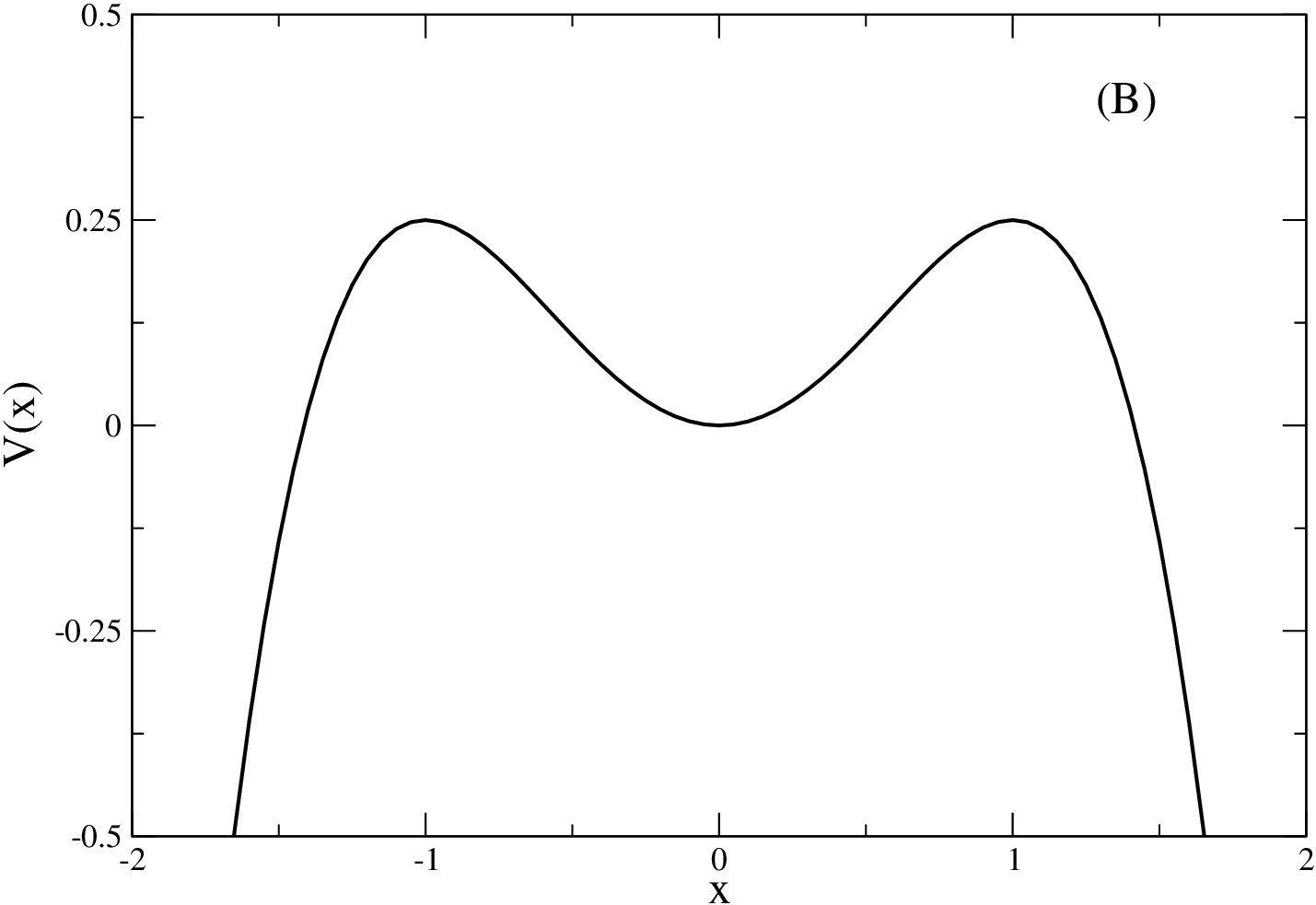}
\caption{Anharmonic potential corresponding to  A) $\omega = \mu =1$ and B) $\omega = 1$ and $\mu = -1$.}
\label{fig4a}
\end{figure}

In Fig.\ref{fig1} we compare the exact frequency, calculated with Eq.~(\ref{periodex}) 
with the frequency obtained with our method (LPLDE),
equation (\ref{omega3duff}), and with the LP method, equation (\ref{omega3duff0}) taking $\lambda=0$,
both  to third order in perturbation theory. We take $\omega = \mu = 1$ (see the left plot of Fig.~\ref{fig4a}) and vary 
the amplitude of the oscillations. 
We observe that our method yields an excellent approximation to the exact result even for large amplitudes, 
where the simple LP approximation fails. 

In Fig.\ref{fig2} we consider the case studied in  Fig.\ref{fig1}, but choosing  $\omega = 1$ and $\mu = - 1$ 
(see the right plot of Fig.~\ref{fig4a}). In this case
the potential has a local minimum in the origin and two maxima, located at $x = \pm 1$. Periodic solutions are supported only
for amplitudes $A < 1$, $A=1$ being a point of (unstable) equilibrium, where the period diverges. Also in this case,
the LPLDE method offers an excellent approximation to the exact result for a large range of amplitudes; as expected, the
approximation is poorer in the region $A \approx 1$, where the point of equilibrium is approached.

\begin{figure}[t]
\includegraphics[width=11cm]{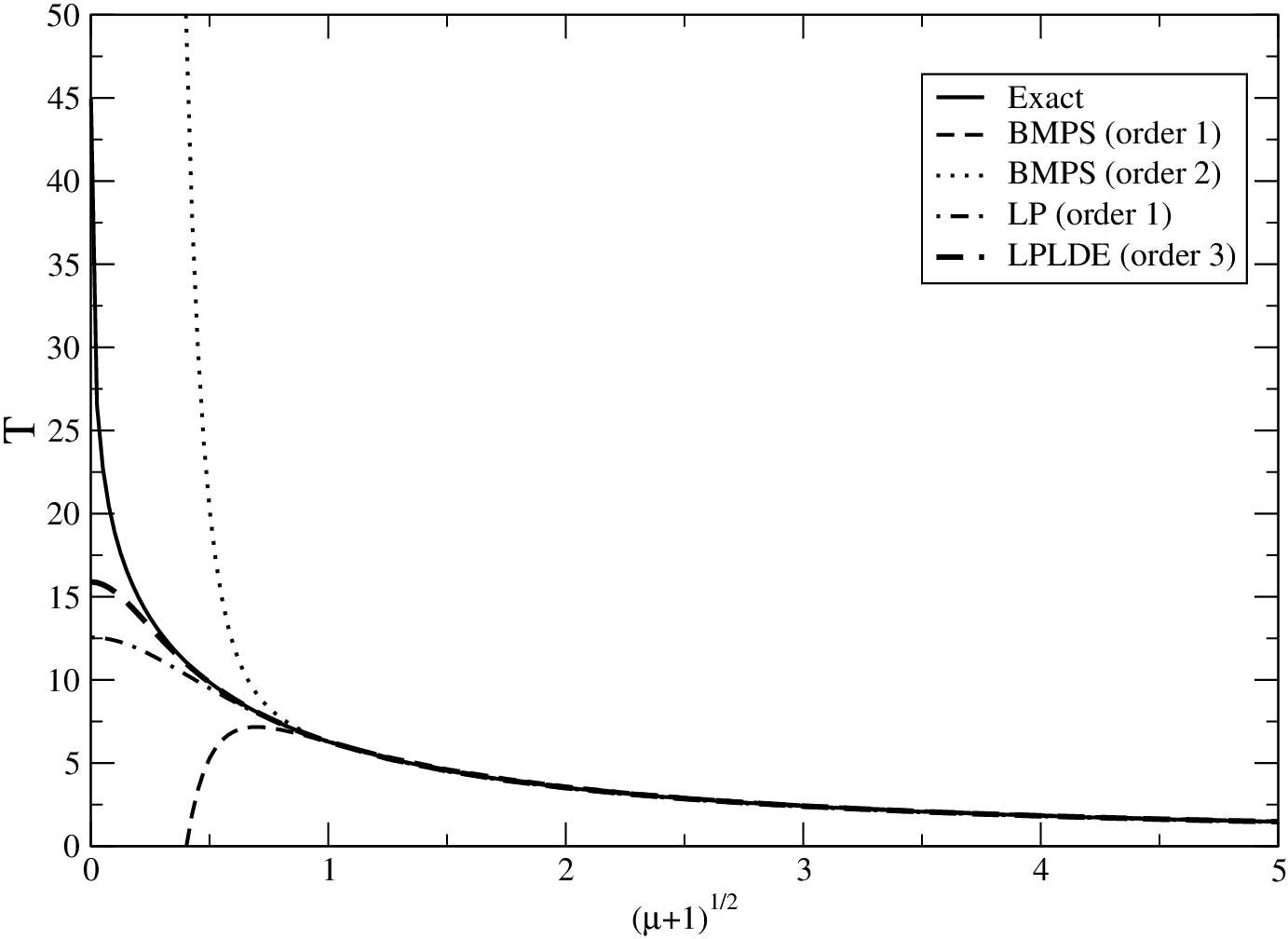}
\caption{Period of the anharmonic oscillator. The curves labeled with BMPS refer to the formulas of  \cite{BMPS89}.}
\label{fig3}
\end{figure}

In Fig.\ref{fig3} we compare the period obtained with our method to the exact period of Eq.~(\ref{periodex}) and
to the one obtained with the formulae of \cite{BMPS89}, which are obtained by applying the nonlinear delta expansion.
Our method provides an excellent approximation to the exact period over a wide range of 
the parameter $\mu$, which controls the 
nonlinearity. The plots are obtained assuming $\omega = 1$ and the boundary conditions $x(0) = 1$ and $\dot{x}(0) = 0$.
The formulae of \cite{BMPS89}  behave badly in the region $\mu < 0$, which corresponds to a potential well of
finite depth centered around $x = 0$, and yield a precision comparable to the one achieved with our method for $\mu >0$. 
Corresponding to the value $\mu = 0$ the oscillator is in a position of (unstable) equilibrium 
and the exact period diverges. Notice that for large values of $\mu$ all the methods seem to give a good approximation
to the exact solution, including the LP method (to first order), which (to third order) was behaving poorly in the case 
previously studied. 
Unfortunately the equations of \cite{BMPS89} are not suitable to be analyzed as in Fig. \ref{fig1}, and thus we cannot fully test
the efficiency of this method.

\begin{figure}[t]
\includegraphics[width=11cm]{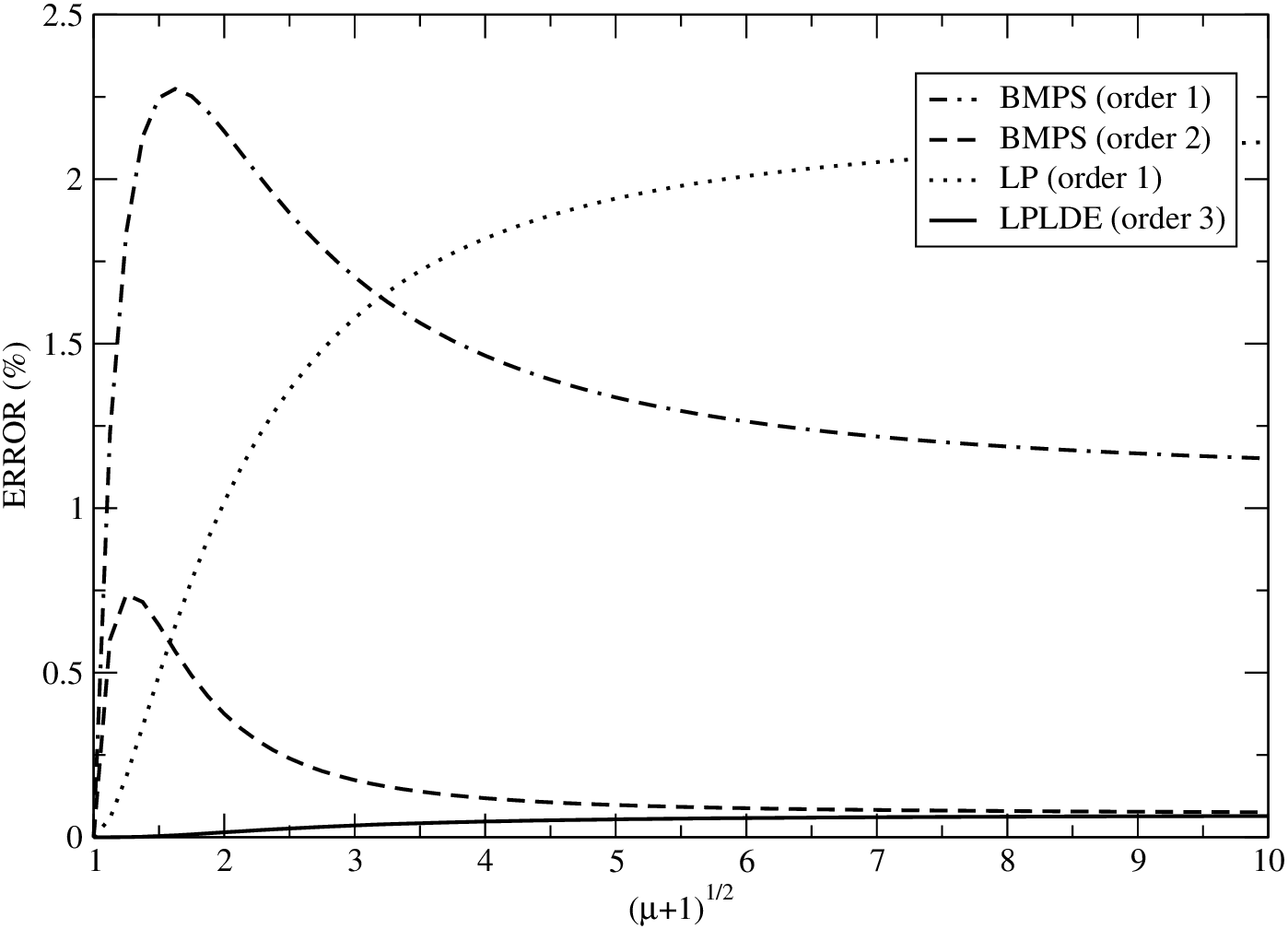}
\caption{Error corresponding to the different approaches for the case studied in Fig. \ref{fig3}. 
The curves labeled with BMPS refer to the formulas of  \cite{BMPS89}.}
\label{fig4}
\end{figure}

In Fig.\ref{fig4} we plot the relative error corresponding to the different approximations for $\mu >0$. Our method
to third order in perturbation theory yields an error typically smaller than the errors of the other methods and with a 
magnitude of about $0.1 \ \%$.

\section{The nonlinear pendulum}
\label{sec:NP}
We now apply the improved method to the nonlinear pendulum. The steps are exactly the same as
before and we proceed to outline them.
First, consider the equation for the nonlinear pendulum
\beq
\frac{d^2\theta}{dt^2} + \omega^2 \  \sin\theta = 0 \, ,
\label{NLP1}
\eeq
where $\omega^2 = g/l$ is the natural frequency of the small oscillations of the pendulum.
Following the Lindstedt-Poincar\'e method, we introduce a scaled time $\tau = \Omega \ t$ and write 
the equation as
\beq
\Omega^2 \ \frac{d^2\theta}{d\tau^2} + \omega^2  \sin\theta = 0 \ ,
\eeq
where $\Omega = 2\pi/T$ is the (unknown) frequency and $T$ is the period of the oscillations.
As discussed in the case of the anharmonic oscillator, we can apply the Linear Delta Expansion to the problem 
by modifying the above equation and writing it as:
\beq
\Omega^2 \ \frac{d^2\theta}{d\tau^2} + \lambda^2 \ \theta = \delta \left[ -\omega^2  \sin\theta
+ \lambda^2 \ \theta\right] \equiv \delta \ f(\theta) \ ,
\label{NLP2}
\eeq 
where $\lambda$ is an arbitrary parameter, with the dimension of frequency. In what follows we use the same 
procedure previously outlined for the anharmonic oscillator, with a few technical differences due to the more
difficult nature of the present problem. 

We will expand the angle and the frequency as
\beq
\theta(\tau) &=& \sum_{n=0}^\infty \ \delta^n \ \theta_n(\tau) \ \ \ , \ \ \
\Omega^2 = \sum_{n=0}^\infty \ \delta^n \ \alpha_n  \nonumber \ .
\eeq

We will solve Eq.~(\ref{NLP2}) subject to the boundary condition $\theta(0) = A$ and $\dot{\theta}(0) = 0$, i.e.
\beq
\theta_0(0)= A  \ \ \ , \ \ \ \theta_{j>0}(0) = 0 \ \ \ , \ \ \ \dot{\theta}_j=0 \ .
\eeq

\subsubsection{Zeroth Order}

To zeroth order the equation for the pendulum reads
\beq
\alpha_0 \ \frac{d^2\theta_0}{d\tau^2} + \lambda^2 \ \theta_0 = 0 \ ,
\eeq
and we obtain the solution
\beq
\theta_0(\tau) = A \ \cos\tau \ ,
\eeq
describing a simple oscillatory motion with (scaled) period $2 \pi$.
The zeroth order frequency is therefore given by
\beq
\alpha_0 = \lambda^2 \, .
\eeq

\subsubsection{First Order}

To first order we obtain the differential equation:
\beq
\alpha_0 \ \frac{d^2\theta_1}{d\tau^2} + \lambda^2 \ \theta_1 &=& S_1(\tau) \ ,
\label{po1:1}
\eeq
where we have defined the source term:
\beq
S_1(\tau) &\equiv& - \alpha_1 \ \frac{d^2\theta_0}{d\tau^2} + f(\theta_0) =
A \ \alpha_1 \ \cos\tau + \left[ - \omega^2 \ \sin \left( A \ \cos\tau \right)  + \lambda^2 \ A \ \cos\tau \right] \ .
\eeq

As before, in order to avoid the occurrence of secular terms, 
we need to eliminate contributions proportional to  $\cos\tau$ from 
the source term $S_1(\tau)$ (recall that such a term would yield a resonant 
behavior of the solution $\theta_1(\tau))$. We enforce this condition by requiring that
\beq
\frac{1}{\pi} \ \int_0^{2 \pi} \ d\tau \ S_1(\tau) \ e^{i \ \tau} = 0 \ .
\eeq

As a result of this operation, we are able to fix the coefficient $\alpha_1$:
\beq
\alpha_1 &=& \frac{1}{A} \ \left[ - \lambda^2 \  A +  
\frac{\omega^2}{\pi} \ \int_0^{2 \pi} \ d\tau \ \sin(A \ \cos\tau) \ e^{i \ \tau} \right]
= \frac{1}{A} \ \left[ - \lambda^2 \  A +  \omega^2 \ c_1 \right] \ ,
\eeq
where we have used the following expansion of  $\sin(A \ \cos\tau)$: 
\beq
\sin(A \ \cos\tau) &=& \sum_{j=0}^\infty \ c_{2 j+1} \ \cos \left[(2 j+1)\tau\right]  \ ,
\eeq
and 
\beq
c_{2 j+1} &=& \frac{1}{\pi} \ \int_0^{2\pi} \ d\tau \ e^{i \ (2 j+1)\tau} \ \sin(A \ \cos\tau) 
= 2 \ (-1)^j \ J_{2 j+1}(A) \ .
\eeq

Eq.~(\ref{po1:1})  now reads
\beq
\alpha_0 \ \frac{d^2\theta_1}{d\tau^2} + \lambda^2 \ \theta_1 &=& S_1(\tau) = 
- \omega^2 \ \sum_{j=1}^\infty \ c_{2 j+1} \ \cos \left[(2 j+1)\tau\right] \ ,
\label{po1:2}
\eeq
where the sum starts from $j=1$ because of the vanishing of the term proportional to $\cos\tau$.

We write the solution $\theta_1(\tau)$ as
\beq
\theta_1(\tau) = \sum_{j=0}^\infty \ d^{(1)}_{2 j+1} \ \cos \left[(2 j+1)\tau\right] \ ,
\eeq
where the coefficients are (for $j>1$):
\beq
d^{(1)}_{2 j+1}  &=&   \frac{\omega^2 \ c_{2 j+1}}{4 \  \lambda^2 j \left(j+1\right) } \equiv
\frac{\overline{d}^{(1)}_{2 j+1}}{\lambda^2} \ .
\eeq
In the last equation we have introduced the scale coefficients $\overline{d}^{(1)}_{2 j+1}$, which do not depend upon $\lambda$.
We notice that Eq.~(\ref{po1:2}) cannot be used to determine the coefficient corresponding to $j=0$; in fact, this
coefficient is fixed by the boundary condition:
\beq
\theta_1(0) &=& \sum_{j=0}^\infty \ d^{(1)}_{2 j+1} = 0 \ ,
\eeq
which entails
\beq
d^{(1)}_1 &=& -  \sum_{j=1}^\infty \ d^{(1)}_{2 j+1} =  -  \sum_{j=1}^\infty \   
\frac{\omega^2 \ c_{2 j+1}}{4 \ \lambda^2 \ j \ \left( j+1 \right) } \equiv \frac{\overline{d}^{(1)}_1}{\lambda^2} \ .
\eeq

\subsubsection{Second Order }

To second order we obtain the equation:
\beq
\alpha_0 \ \frac{d^2\theta_2}{d\tau^2} + \lambda^2 \ \theta_2 &=& S_2(\tau) \ ,
\label{po2:1}
\eeq
where we have introduced the source term
\beq
S_2(\tau) &\equiv&  - \alpha_1 \ \frac{d^2\theta_1}{d\tau^2} 
 - \alpha_2 \ \frac{d^2\theta_0}{d\tau^2} + \theta_1(\tau) \ f'(\theta_0) \nonumber \\
&=& - \alpha_1 \ \frac{d^2\theta_1}{d\tau^2} 
 - \alpha_2 \ \frac{d^2\theta_0}{d\tau^2} + \theta_1(\tau) \ \left[- \omega^2 \ \cos\theta_0(\tau)+\lambda^2 \right] \ .
\eeq

We can expand the source term in a series as
\beq
S_2(\tau) &=& \sum_{n=1}^\infty \ s_{2 n+1}^{(2)} \ \cos(2 n+1)\tau \ ,
\eeq
where the coefficients of the expansion are given by
\beq
s^{(2)}_{2 n+1} &=& \frac{1}{\pi} \ \int_0^{2\pi} d\tau \ e^{i \ (2 n+1) \tau} \ S_2(\tau) 
\equiv \frac{\overline{s}^{(2a)}_{2 n+1}}{\lambda^2} + \overline{s}^{(2b)}_{2 n+1} \ .
\eeq
We have introduced the scaled coefficients $\overline{s}^{(2a)}_{2 n+1}$ and  $\overline{s}^{(2b)}_{2 n+1}$, which are independent
of $\lambda$ and read:
\beq
\overline{s}^{(2a)}_{2 n+1} &=&  \frac{\omega^2 \ c_1}{A} \ (2 n +1 )^2 \ \overline{d}^{(1)}_{2 n+1} 
\nonumber \\
&-& \frac{\omega^2}{2} \ \left\{ \sum_{j=n}^\infty \  \overline{d}^{(1)}_{2 j+1} \ \tilde{c}_{2 (j-n)} + 
\sum_{l=n+1}^\infty \   \overline{d}^{(1)}_{2 (l-n-1)+1} \ \tilde{c}_{2 l} + 
\sum_{j=0}^n \ \overline{d}^{(1)}_{2 j+1} \   \tilde{c}_{2 (n-j)}
\right\} \, ,\nonumber \\
\overline{s}^{(2b)}_{2 n+1} &=& - 4 \ n \ (n+1) \ \overline{d}^{(1)}_{2 n+1} \nonumber \ .
\eeq

The coefficients $\tilde{c}_{2 j}$ follow from the expansion of $\cos\left[ A \cos\tau\right]$:
\beq
\cos\left[ A \cos\tau\right] &=& \sum_{j=0}^\infty \ \tilde{c}_{2 j} \ \cos \left[ 2 j \ \tau\right]
\eeq
and read, for $j>0$,
\beq
\tilde{c}_{2 j} &\equiv& \frac{1}{\pi} \ \int_0^{2\pi} \ d\tau \ \cos\left[ A \cos\tau\right]
\ e^{i \  2 j \ \tau} \nonumber \\
&=& 2 \sum_{n=j}^\infty \ (-1)^n \ \left( \frac{A}{2} \right)^{2 n} \ \frac{1}{(n-j)! \ (n+j)!} 
=  2 \ (-1)^j \ J_{2 j}(A) 
\eeq
and, for $j=0$,
\beq
\tilde{c}_0 &=& \cos A - \sum_{j=1}^\infty \ \tilde{c}_{2j} \ .
\eeq

As before we need to eliminate the coefficient $s^{(2)}_{1}$
\beq
s^{(2)}_{1} &=&  \alpha_1 \ d^{(1)}_{1} + \alpha_2 \  A + \lambda^2 \ d^{(1)}_{1} 
- \frac{\omega^2}{2} \ \sum_{j=0}^\infty \ d^{(1)}_{2 j+1} \ \left( \tilde{c}_{2 j} + \tilde{c}_{2 j+2} \right) 
- \frac{\omega^2}{2} \  d^{(1)}_{1} \ \tilde{c}_{0} 
\eeq
and obtain the coefficient $\alpha_2$
\beq
\alpha_2 &=& \frac{1}{A} \ \left\{ - \left( \frac{\omega^2 \ c_1}{A} - \frac{\omega^2 \ \tilde{c}_0}{2}  \right) \  
\frac{\overline{d}^{(1)}_{1}}{\lambda^2} + \frac{\omega^2}{2 \ \lambda^2} \ 
\sum_{j=0}^\infty \ \overline{d}^{(1)}_{2 j+1} \  \left( \tilde{c}_{2 j} + \tilde{c}_{2 j+2} \right) 
\right\} \equiv \frac{\overline{\alpha}_2}{\lambda^2} \ ,
\eeq
where $\overline{\alpha}_2 = \lambda^2 \alpha_2$ is a scaled coefficient, independent of $\lambda$.

We are therefore able to find the solution of Eq.~(\ref{po2:1}) 
\beq
\theta_2(\tau) = \sum_{j=0}^\infty \ d^{(2)}_{2 j+1} \ \cos \left[(2 j+1)\tau\right] \ ,
\eeq
with the coefficients, for $j \neq 1$ 
\beq
d_{2j+1}^{(2)} \equiv \frac{\overline{d}^{(2a)}_{2 j +1}}{\lambda^4} +  \frac{\overline{d}^{(2b)}_{2 j +1}}{\lambda^2} \ ,
\eeq
expressed in terms of the $\lambda$-independent terms:
\beq
\overline{d}^{(2a)}_{2 j +1} &=&  - \frac{\overline{s}^{(2a)}_{2 j+1}}{4 \ j \ (j+1)} \, , \nonumber \\
\overline{d}^{(2b)}_{2 j +1} &=&  - \frac{\overline{s}^{(2b)}_{2 j+1}}{4 \ j \ (j+1)} 
= \overline{d}^{(1)}_{2 j +1} \nonumber \ .
\eeq

As before the $j=0$ coefficient is not fixed by the equation and needs to be determined by enforcing the boundary
condition $\theta_2(0) = 0$. We obtain:
\beq
d_1^{(2)} &=& -  \sum_{j=1}^\infty \ d^{(2)}_{2 j+1} =  
\sum_{j=1}^\infty \ \frac{s_{2j+1}^{(2)}}{4 \ \lambda^2 \ j \ (j+1)} \, .
\eeq

\subsubsection{Third Order }

To third order we obtain the equation
\beq
\alpha_0 \ \frac{d^2\theta_3}{d\tau^2} + \lambda^2 \ \theta_3 &=& S_3(\tau) \, ,
\eeq
where the source term $S_3(\tau)$ is 
\beq
S_3(\tau) &\equiv& - \alpha_1 \ \frac{d^2\theta_2}{d\tau^2} 
 - \alpha_2 \ \frac{d^2\theta_1}{d\tau^2}   - \alpha_3 \ \frac{d^2\theta_0}{d\tau^2} 
+ \left[\theta_2(\tau) \ f'(\theta_0) + \frac{\theta_1^2(\tau)}{2} \ f''(\theta_0) \right] \ .
\eeq

Once again it is useful to expand the source term in a series as
\beq
S_3(\tau) &=& \sum_{n=0}^\infty \ s_{2 n+1}^{(3)} \ \cos(2 n+1)\tau \, ,
\eeq
where the coefficients of the expansion are given by
\beq \label{s3n}
s^{(3)}_{2 n+1} &=& \frac{1}{\pi} \ \int_0^{2\pi} d\tau \ e^{i \ (2 n+1) \tau} \ S_3(\tau) = 
\frac{\overline{s}^{(3a)}_{2 n+1}}{\lambda^4} + \frac{\overline{s}^{(3b)}_{2 n+1}}{\lambda^2} + 
\overline{s}^{(3c)}_{2 n+1} 
\eeq
and $\overline{s}^{(3a,b,c)}$ are independent of $\lambda$. 
A lengthy calculation  allows to find the expressions for these coefficients, which can be found in Appendix A.
Here we only write the coefficient of the term  $\cos\tau$, corresponding to $n=0$:
\beq
s^{(3)}_{1} &=& \alpha_1 \ \ d_{1}^{(2)} + \alpha_2 \  d_{1}^{(1)} + \alpha_3 \ A \  
- \frac{\omega^2}{2} \ \sum_{j=0}^\infty  \ \tilde{c}_{2 j} \ d^{(2)}_{2 j+1} 
 - \frac{\omega^2}{2} \ \sum_{l=1}^\infty  \ \tilde{c}_{2 l} \ d^{(2)}_{2 l-1} \nonumber \\ 
&-& \frac{\omega^2}{2} \  \tilde{c}_{0} \ d^{(2)}_{1} +  \lambda^2 \ d_{1}^{(2)} \nonumber \\ 
&+& \frac{\omega^2}{8} \ \sum_{m=0}^{\infty} \sum_{j=m+1}^{\infty} \ c_{2 (j-m-1) +1} \ d_{2 m+1}^{(1)} \ d_{2 j+1}^{(1)}
+ \frac{\omega^2}{8} \ \sum_{j=0}^{\infty} \sum_{m=0}^{j} \ c_{2 (-m+j) +1} \ d_{2 m+1}^{(1)} \ d_{2 j+1}^{(1)} \nonumber \\ 
&+&\frac{\omega^2}{8} \ \sum_{m=j+1}^{\infty} \sum_{j=0}^{\infty} \ c_{2 (m-j-1) +1} \ d_{2 j+1}^{(1)} \ d_{2 m+1}^{(1)}
+ \frac{\omega^2}{8} \ \sum_{m=0}^{\infty} \sum_{j=0}^{m} \ c_{2 (m-j) +1} \ d_{2 j+1}^{(1)} \ d_{2 m+1}^{(1)} \nonumber \\ 
&+& \frac{\omega^2}{8} \ \sum_{m=0}^{\infty} \sum_{j=0}^{\infty} \ c_{2 (m+j) +1} \ d_{2 j+1}^{(1)} \ d_{2 m+1}^{(1)}
+ \frac{\omega^2}{8} \ \sum_{m=0}^{\infty} \sum_{j=0}^{\infty} \ c_{2 (m+j+1) +1} \ d_{2 j+1}^{(1)} \ d_{2 m+1}^{(1)} \ .
\eeq

The coefficient $\alpha_3$ is fixed by requiring that $s^{(3)}_{1}$ vanish:
\beq
\alpha_3 &=& \frac{\alpha_{3a}}{\lambda^4} + \frac{\alpha_{3b}}{\lambda^2}  \ ,
\eeq
where
\beq
\alpha_{3a} &=&-\frac{\omega^2}{A} \ \left\{ \left( \frac{c_1}{A} - \frac{\tilde{c}_0}{2} \right) \ 
\overline{d}^{(2a)}_{1} + \overline{\alpha}_2 \ \overline{d}_1^{(1)}  
- \frac{1}{2} \ \sum_{j=0}^\infty \ \left( \tilde{c}_{2j} + \tilde{c}_{2 j+2} \right) \ 
\overline{d}^{(2a)}_{2 j +1} \right.  \nonumber \\
&+& \left. \frac{1}{8} \ \sum_{m=0}^\infty \ \left[ \sum_{j=m+1}^\infty \ \left( 2 \ c_{2 (j-m-1)+1} + 
c_{2 (m+j)+1} + c_{2 (m+j)+3} \right) \ \overline{d}_{2 j+1}^{(1)} \ \overline{d}_{2 m+1}^{(1)}  \right. \right. \nonumber \\
&+& \left. \left. \sum_{j=0}^m \ \left( 2 \ c_{2 (m-j)+1} + 
c_{2 (m+j)+1} + c_{2 (m+j)+3} \right) \ \overline{d}_{2 j+1}^{(1)} \ \overline{d}_{2 m+1}^{(1)}  \right]
\right\} \, , \nonumber \\
\alpha_{3b} &=& -\frac{\omega^2}{A} \ \left\{ \left( \frac{c_1}{A} - \frac{\tilde{c}_0}{2} \right) \ 
\overline{d}^{(2b)}_{1} - \frac{1}{2} \ \sum_{j=0}^\infty \ \left( \tilde{c}_{2j} + \tilde{c}_{2 j+2} \right) \ 
\overline{d}^{(2b)}_{2 j +1} \right\}  = \overline{\alpha}_2 \, .
\eeq

To this order the squared frequency reads:
\beq
\Omega^2 &=& \alpha_{1a} + 2 \ \frac{\overline{\alpha}_{2}}{\lambda^2} + \frac{\alpha_{3a}}{\lambda^4} \nonumber \ .
\eeq

The ``principle of minimal sensitivity'' yields  the solution
\beq
\lambda^2 &=& - \frac{\alpha_{3a}}{\overline{\alpha}_{2}} 
\eeq
and a corresponding value of $\Omega^2$:
\beq
\Omega^2 &=&  \alpha_{1a} - \frac{\overline{\alpha}_2^2}{\alpha_{3a}} \ .
\eeq

\begin{figure}[t]
\includegraphics[width=11cm]{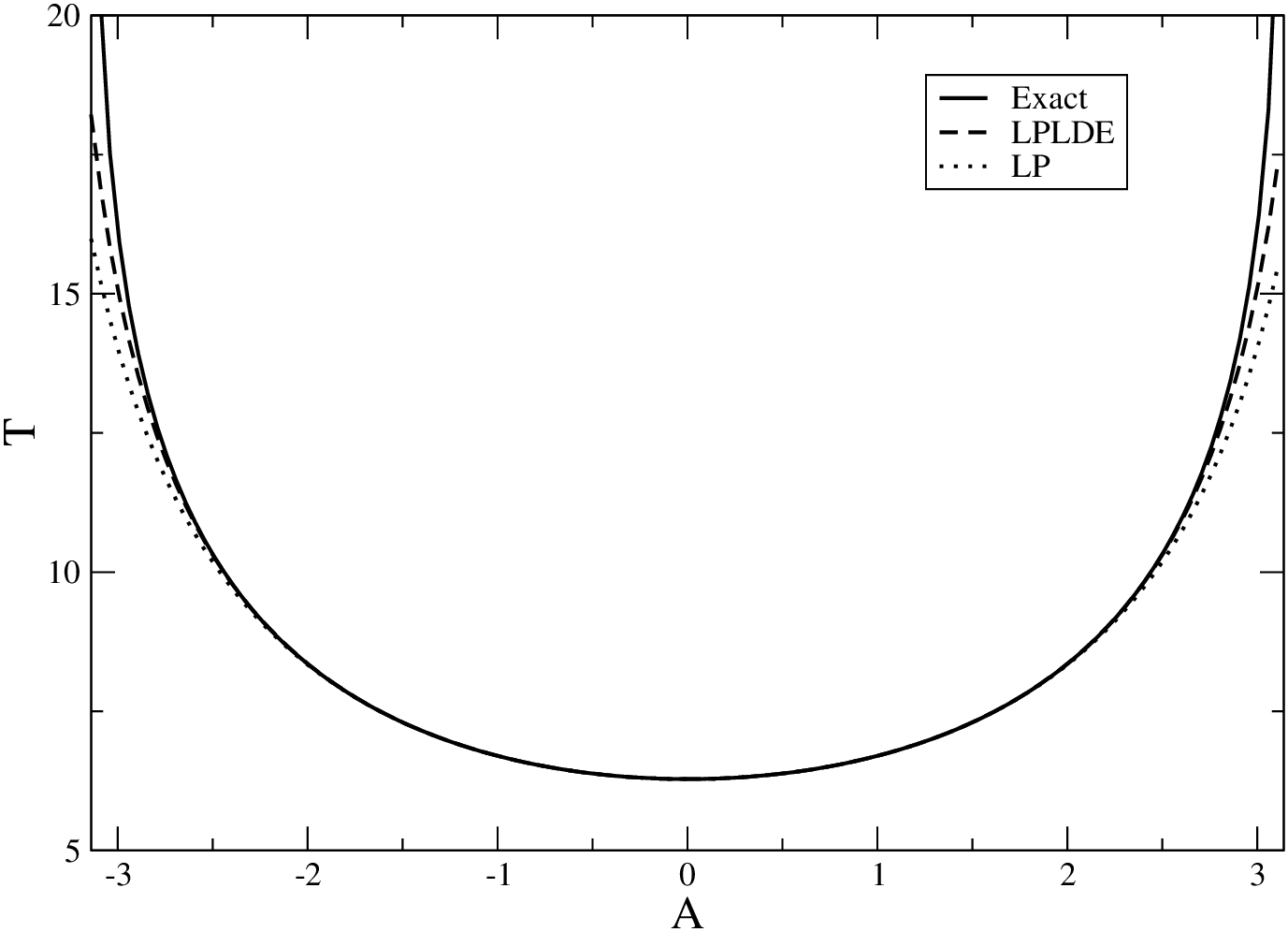}
\caption{Period of the nonlinear pendulum as a function of the amplitude. We assume $\omega = \sqrt{g/l} = 1$. }
\label{fig5}
\end{figure}

In Fig. \ref{fig5} we plot the period of the nonlinear pendulum as a function of the amplitude, as obtained in the
LPLDE and LP approximations, and compare the results with the exact period. We assume $\omega = 1$ and use the formulae
given above truncating the infinite series to a maximum value $j_{max}= 5$. As it can be seen from the Figure, the LPLDE 
approximation is in excellent agreement with the exact result, up to very large amplitudes. $A = \pm \pi$ corresponds to 
an unstable point of equilibrium, for which the exact period diverges.

\section{Conclusions}
\label{sec:conclusions}
We have presented a method for the solution of nonlinear problems which are conservative and periodic.
It is based on the application
of the Linear Delta Expansion to the Lindstedt-Poincar\'e method. We applied it to two problems: the
Duffing Equation and the nonlinear pendulum. In the case of the Duffing equation we find that the 
new model converges faster and with greater accuracy than the simple LP method. Also, by comparing it
with methods based on the perturbative $\delta$ expansion, we show that our solution not only converges faster and
more accurately, but it also works for a much wider range of parameters, including the case in which the
nonlinear coupling $\mu$ is negative. 
In a similar fashion, we show that the method works remarkably well for the solution of the nonlinear pendulum, for which 
the method is implented without performing any Taylor expansion of the potential.

We are currently working on the extension of the present method to quantum systems and multiple scale analysis \cite{WP}.

\section*{Acknowledgements}

The Authors wish to thank Prof. J.D.Walecka for his useful comments.
They also acknowledge the support of the ``Fondo Ram\'on Alvarez-Buylla'' of the University of Colima and of Conacyt 
in the completion of this work.


\appendix

\section{Coefficients}
In this Appendix we present the computation of the coefficients of $s^{(3)}_{2n+1}$ in Eq.~(\ref{s3n}).
Let us rewrite Eq.~(\ref{s3n}) in the following form:
\beq
s^{(3)}_{2 n+1} &\equiv& I^{(A)}_{2 n+1} + I^{(B)}_{2 n+1} + I^{(C)}_{2 n+1} + I^{(D)}_{2 n+1} \, .
\eeq
We now proceed to compute each of these terms:
\begin{itemize}

\item \underline{$I^{(A)}_{2n+1}$}

\beq
I^{(A)}_{2 n+1} 
&=&  \left(\frac{\omega^2 \ c_1}{A} - \lambda^2 \right) \ (2 n+1)^2 \ 
\left[ \frac{\overline{d}^{(2a)}_{2 n +1}}{\lambda^4} +  \frac{\overline{d}^{(2b)}_{2 n +1}}{\lambda^2} \right]
\\ \nonumber
&+& \frac{\overline{\alpha}_2}{\lambda^2} \  (2 n+1)^2 \  \frac{\overline{d}_{2 n+1}^{(1)}}{\lambda^2} 
+ \alpha_3 \ A \ \delta_{n0} \nonumber \\
&\equiv& \frac{i_a^{(1)}}{\lambda^4} + \frac{i_a^{(2)}}{\lambda^2} +  i_a^{(3)}
\eeq

\item \underline{$I^{(B)}_{2 n+1}$}

\beq
 I^{(B)}_{2 n+1}  &=& - \frac{\omega^2}{2} \ \sum_{l=0}^\infty  \ \sum_{j=0}^\infty \ \tilde{c}_{2 l} \ d^{(2)}_{2 j+1} \ 
\left[ \delta_{n+l-j,0} + \delta_{n+1-l+j,0} + \delta_{l+j-n,0} + \delta_{n+l+j+1,0} \right]  \nonumber \\
&\equiv& \frac{i_b^{(1)}}{\lambda^4} + \frac{i_b^{(2)}}{\lambda^2}
\eeq
The four different integrals become:
\begin{itemize}
\item a)
\beq
 I^{(B1)}_{2 n+1}  &=& - \frac{\omega^2}{2} \ \sum_{j=n}^\infty  \ \tilde{c}_{2 (j-n)} \ d^{(2)}_{2 j+1} 
\eeq

\item b)
\beq
 I^{(B2)}_{2 n+1}  &=&  - \frac{\omega^2}{2} \ \sum_{l=n+1}^\infty  \ \tilde{c}_{2 l} \ d^{(2)}_{2 (l-n-1)+1} 
\eeq

\item c)
\beq
 I^{(B3)}_{2 n+1}  &=&  - \frac{\omega^2}{2} \ \sum_{l=0}^n  \ \tilde{c}_{2 l} \ d^{(2)}_{2 (n-l)+1} 
\eeq

\item d)
\beq
 I^{(B4)}_{2 n+1}  &=& 0 
\eeq

\end{itemize}

Therefore we finally have that
\beq
I^{(B)}_{2 n+1} &=& - \frac{\omega^2}{2} \ \left[ \sum_{j=n}^\infty  \ \tilde{c}_{2 (j-n)} \ d^{(2)}_{2 j+1}  
        + \sum_{l=n+1}^\infty  \ \tilde{c}_{2 l} \ d^{(2)}_{2 (l-n-1)+1}  
        + \sum_{l=0}^n  \ \tilde{c}_{2 l} \ d^{(2)}_{2 (n-l)+1} \right] \ .
\eeq

\item \underline{$I^{(C)}_{2n+1}$}

\beq
I^{(C)}_{2n+1} &=&  \frac{1}{\pi} \ \int_0^{2\pi} d\tau \ e^{i \ (2 n+1) \tau} \ \lambda^2 \
\left(\sum_{j=0}^\infty \ d_{2j+1}^{(2)} \ \cos (2 j+1)\tau \right)  = \lambda^2 \ d_{2 n+1}^{(2)} \nonumber \\
&\equiv& \frac{i_c^{(1)}}{\lambda^2} + i_c^{(2)}
\eeq

\item \underline{$I^{(D)}_{2n+1}$}
\beq
I^{(D)}_{2n+1} 
 &=&  \frac{1}{\pi} \ \int_0^{2\pi} d\tau \ e^{i \ (2 n+1) \tau} \ \frac{\omega^2}{2} \ \sin A\cos\tau \
\left(\sum_{j=0}^\infty \ d_{2j+1}^{(1)} \ \cos (2 j+1)\tau \right)^2 \nonumber \\
 &=&  \frac{1}{\pi} \ \int_0^{2\pi} d\tau \ e^{i \ (2 n+1) \tau} \ \frac{\omega^2}{2} \ \sum_{l=0}^\infty c_{2 l+1} \ 
\cos (2 l+1)\tau \\ \nonumber 
 & & \times \sum_{m=0}^\infty \ d_{2m+1}^{(1)} \ \cos (2 m+1)\tau \sum_{j=0}^\infty \ d_{2j+1}^{(1)} \ \cos (2 j+1)\tau 
\eeq

We need to calculate the following integral:
\beq
{\cal I} &=& \frac{1}{\pi} \ \int_0^{2 \pi} \ d\tau \ e^{i (2 n+1) \tau} \ 
\cos\left[(2 j+1)\tau\right] \ \cos\left[(2 l +1)\tau\right]
\ \cos\left[(2 m + 1) \tau\right] \, .
\eeq
Using the relation
\beq
{\cal C} &=& \cos\left[(2 j+1)\tau\right] \ \cos\left[(2 l +1)\tau\right] \ \cos\left[(2 m + 1) \tau\right]  \nonumber \\
&=& \frac{1}{4} \ \left[ \cos (2(l+m+j)+3)\tau + \cos (2(l+m-j)+1)\tau \right. \nonumber \\
&+& \left. \cos (2(l-m+j)+1)\tau + \cos (2(l-m-j)-1)\tau 
\right] \, , 
\eeq
one obtains
\beq
{\cal I} &=& \frac{1}{\pi} \ \int_0^{2 \pi} \ d\tau \ e^{i (2 n+1) \tau} \ {\cal C} = \frac{1}{4} 
\ \left\{ \delta_{2 (n+l+m+j)+4,0} + \delta_{2 (n-l-m-j)-2,0} \right. \nonumber \\
&+& \left.
\delta_{2 (n+l+m-j)+2,0} + \delta_{2 (n-l-m+j),0}+\delta_{2 (n+l-m+j)+2,0} + \delta_{2 (n-l+m-j),0} \right. \nonumber \\
&+& \left.
\delta_{2 (n+l-m-j),0} + \delta_{2 (n-l+m+j)+2,0}    
\right\} \, ,
\eeq
and finally 
\beq
I^{(D)}_{2n+1}  
&=&  \frac{\omega^2}{8} \ \sum_{l=0}^\infty \sum_{m=0}^\infty \sum_{j=0}^\infty \ c_{2 l +1} \ d_{2 m+1}^{(1)} \  
d_{2 j+1}^{(1)} \ \left\{ \delta_{2 (n+l+m+j)+4,0} + \delta_{2 (n-l-m-j)-2,0} \right. \nonumber \\
&+& \left.
\delta_{2 (n+l+m-j)+2,0} + \delta_{2 (n-l-m+j),0}+\delta_{2 (n+l-m+j)+2,0} + \delta_{2 (n-l+m-j),0} \right. \nonumber \\
&+& \left. \delta_{2 (n+l-m-j),0} + \delta_{2 (n-l+m+j)+2,0}   \right\} = \frac{i_d}{\lambda^4} \, .
\eeq
We are then left with 8 integrals that can be evaluated in the following way (we call them $I_D^{(i)}$):
\begin{itemize}

\item i)

\beq
I_D^{(1)}  &=&  \frac{\omega^2}{8} \ \sum_{l=0}^\infty \sum_{m=0}^\infty \sum_{j=0}^\infty \ c_{2 l +1} \ d_{2 m+1}^{(1)} \  
d_{2 j+1}^{(1)} \ \delta_{2 (n+l+m+j)+4,0} = 0
\eeq

\item ii)

\beq
I_D^{(2)}  &=&  \frac{\omega^2}{8} \ \sum_{l=0}^\infty \sum_{m=0}^\infty \sum_{j=0}^\infty \ c_{2 l +1} \ d_{2 m+1}^{(1)} \  
d_{2 j+1}^{(1)} \ \delta_{2 (n-l-m-j)-2,0} \nonumber \\
&=& \frac{\omega^2}{8} \ \sum_{m=0}^{n-j-1} \sum_{j=0}^{n-1} \ c_{2 (n-m-j-1) +1} \ d_{2 m+1}^{(1)} \ d_{2 j+1}^{(1)}
\eeq

\item iii)

\beq
I_D^{(3)}  &=&  \frac{\omega^2}{8} \ \sum_{l=0}^\infty \sum_{m=0}^\infty \sum_{j=0}^\infty \ c_{2 l +1} \ d_{2 m+1}^{(1)} \  
d_{2 j+1}^{(1)} \ \delta_{2 (n+l+m-j)+2,0} \nonumber \\
&=& \frac{\omega^2}{8} \ \sum_{m=0}^{\infty} \sum_{j=n+m+1}^{\infty} \ c_{2 (j-n-m-1) +1} \ d_{2 m+1}^{(1)} \ d_{2 j+1}^{(1)}
\eeq

\item iv)

\beq
I_D^{(4)}  &=&  \frac{\omega^2}{8} \ \sum_{l=0}^\infty \sum_{m=0}^\infty \sum_{j=0}^\infty \ c_{2 l +1} \ d_{2 m+1}^{(1)} \  
d_{2 j+1}^{(1)} \ \delta_{2 (n-l-m+j),0} \nonumber \\
&=& \frac{\omega^2}{8} \ \sum_{j=0}^{\infty} \sum_{m=0}^{n+j} \ c_{2 (n-m+j) +1} \ d_{2 m+1}^{(1)} \ d_{2 j+1}^{(1)}
\eeq

\item v)

\beq
I_D^{(5)}  &=&  \frac{\omega^2}{8} \ \sum_{l=0}^\infty \sum_{m=0}^\infty \sum_{j=0}^\infty \ c_{2 l +1} \ d_{2 m+1}^{(1)} \  
d_{2 j+1}^{(1)} \ \delta_{2 (n+l-m+j)+2,0} \nonumber \\
&=& \frac{\omega^2}{8} \ \sum_{m=n+j+1}^{\infty} \sum_{j=0}^{\infty} \ c_{2 (m-n-j-1) +1} \ d_{2 j+1}^{(1)} \ d_{2 m+1}^{(1)}
\eeq

\item vi)

\beq
I_D^{(6)}  &=&  \frac{\omega^2}{8} \ \sum_{l=0}^\infty \sum_{m=0}^\infty \sum_{j=0}^\infty \ c_{2 l +1} \ d_{2 m+1}^{(1)} \  
d_{2 j+1}^{(1)} \ \delta_{2 (n-l+m-j),0} \nonumber \\
&=& \frac{\omega^2}{8} \ \sum_{m=0}^{\infty} \sum_{j=0}^{m+n} \ c_{2 (m+n-j) +1} \ d_{2 j+1}^{(1)} \ d_{2 m+1}^{(1)}
\eeq

\item vii)

\beq
I_D^{(7)}  &=&  \frac{\omega^2}{8} \ \sum_{l=0}^\infty \sum_{m=0}^\infty \sum_{j=0}^\infty \ c_{2 l +1} \ d_{2 m+1}^{(1)} \  
d_{2 j+1}^{(1)} \ \delta_{2 (n+l-m-j),0} \nonumber \\
&=& \frac{\omega^2}{8} \ \sum_{m=max(0,n-j)}^{\infty} \sum_{j=0}^{\infty} \ c_{2 (m+j-n) +1} \ d_{2 j+1}^{(1)} \ d_{2 m+1}^{(1)}
\eeq

\item viii)

\beq
I_D^{(8)}  &=&  \frac{\omega^2}{8} \ \sum_{l=0}^\infty \sum_{m=0}^\infty \sum_{j=0}^\infty \ c_{2 l +1} \ d_{2 m+1}^{(1)} \  
d_{2 j+1}^{(1)} \  \delta_{2 (n-l+m+j)+2,0}   \nonumber \\
&=& \frac{\omega^2}{8} \ \sum_{m=0}^{\infty} \sum_{j=0}^{\infty} \ c_{2 (n+m+j+1) +1} \ d_{2 j+1}^{(1)} \ d_{2 m+1}^{(1)}
\eeq
\end{itemize} 
The final expression is:
\beq
I^{(D)}_{2n+1} 
&=&  \frac{\omega^2}{8} \ \left\{\sum_{m=0}^{n-j-1} \sum_{j=0}^{n-1} \ c_{2 (n-m-j-1) +1} \ d_{2 m+1}^{(1)} \ d_{2 j+1}^{(1)}
+ \sum_{m=0}^{\infty} \sum_{j=n+m+1}^{\infty} \ c_{2 (j-n-m-1) +1} \ d_{2 m+1}^{(1)} \ d_{2 j+1}^{(1)} \right. \nonumber \\
&+& \left. \sum_{j=0}^{\infty} \sum_{m=0}^{n+j} \ c_{2 (n-m+j) +1} \ d_{2 m+1}^{(1)} \ d_{2 j+1}^{(1)}
+ \sum_{m=n+j+1}^{\infty} \sum_{j=0}^{\infty} \ c_{2 (m-n-j-1) +1} \ d_{2 j+1}^{(1)} \ d_{2 m+1}^{(1)} \right. \nonumber \\
&+& \left. \sum_{m=0}^{\infty} \sum_{j=0}^{m+n} \ c_{2 (m+n-j) +1} \ d_{2 j+1}^{(1)} \ d_{2 m+1}^{(1)}
+ \sum_{m=max(0,n-j)}^{\infty} \sum_{j=0}^{\infty} \ 
c_{2 (m+j-n) +1} \ d_{2 j+1}^{(1)} \ d_{2 m+1}^{(1)} \right. \nonumber \\
&+&\left.  \sum_{m=0}^{\infty} \sum_{j=0}^{\infty} \ c_{2 (n+m+j+1) +1} \ d_{2 j+1}^{(1)} \ d_{2 m+1}^{(1)}
\right\}
\eeq

\end{itemize}

\newpage


\end{document}